\newcommand{\K}{\ensuremath{\textrm{ K}}}
\newcommand{\pc}{\ensuremath{\textrm{ pc}}}
\newcommand{\kpc}{\ensuremath{\textrm{ kpc}}}
\newcommand{\cm}{\ensuremath{\textrm{ cm}}}

\newcommand{\kms}{\ensuremath{\textrm{ km s}^{-1}}}
\newcommand{\MHz}{\ensuremath{\textrm{ MHz}}}
\newcommand{\radmsq}{\ensuremath{\textrm{ rad m}^{-2}}}
\newcommand{\uG}{\ensuremath{\, \mu \textrm{G}}}
\newcommand{\cucm}{\ensuremath{\cm^{-3}}}
\newcommand{\PI}{\ensuremath{\textrm{ PI}}}
\renewcommand{\vec}{\mathbf}

\documentclass[twocolumn]{aastex631}
\usepackage{units}

\newcommand{\RM}{\ensuremath{\mathrm{RM}}}
\newcommand{\vlsr}{\ensuremath{V_\mathrm{LSR}}}

\begin{document}

\title{What Rogue Hydrogen Clouds tell us about Galactic Magnetic Fields}

\newcommand{\nrcaffil}{\affiliation{National Research Council Canada, Herzberg Astronomy and Astrophysics, Dominion Radio Astrophysical Observatory, Kaleden, BC Canada}}

\author{Bailey M. Forster}
\affiliation{Dept. of Physics \& Astronomy, Brandon University}
\nrcaffil
\affiliation{Dept. of Physics, McGill University}
\author{Tyler J. Foster}
\email{FosterT@BrandonU.CA}
\correspondingauthor{Tyler J. Foster}
\affiliation{Dept. of Physics \& Astronomy, Brandon University}
\author{Roland Kothes}
 \nrcaffil
\author{Alex S. Hill}
\affiliation{Dept. of Computer Science, Math, Physics, \& Statistics, University of British Columbia, Kelowna, BC Canada}
\nrcaffil
\author{Jo-Anne C. Brown}
\affiliation{Dept. of Physics \& Astronomy, University of Calgary}

\begin{abstract}
Observations of the Milky Way and external galaxies support the idea that large-scale magnetic fields are concentrated in galactic disks, with halo magnetic fields at least an order of magnitude weaker. However, very little is known about the transition between the two.  We present the discovery of linearly polarized radio emission at the interface between interacting shells of gas within a well-known grouping of high-velocity clouds (HVCs), the Anti-Center Shell. Faraday rotation of diffuse emission and of background extragalactic compact sources demonstrates an enhancement of the field at the interface. This is the clearest observed example of an HVC altering the large-scale magnetic field at the disk-halo interface and is the first image of magnetic field effects in an HVC. These results demonstrate the possibility of future three-dimensional reconstruction of the Galactic magnetic field and showcase the versatility of the Synthesis Telescope at the Dominion Radio Astrophysical Observatory as one of the few existing telescopes which can exploit this new method of probing Galactic magnetism. 
\end{abstract}

\section{Introduction}
The interstellar medium (ISM) of our Galaxy has three main constituents: gas and dust in different phases, relativistic cosmic rays, and magnetic fields \citep{Ferriere:2001}. Cosmic rays spiral around the magnetic field threaded through the gas, generating linearly polarized synchrotron emission. Magnetic fields cause Faraday rotation of linearly polarized radiation, allowing us to infer the properties of the magnetic field along the line of sight.

Faraday rotation observations from compact extragalactic sources \citep{SNK80,brow01,brown07,tayl09,vaneck11,Hutschenreuter:2022,Anderson:Heald:2021,Anderson:McClure-Griffiths:2024} and diffuse emission \citep{land10,Hill:2017,Ma:Mao:2020,Dickey:2022} show evidence that the magnetic field in the disk has a spiral pattern with possible field reversals \citep{han06,Ordog:2017}. These data have been used to constrain models of the large-scale Galactic magnetic field \citep{sun08,jansson12}. In the halo, \citet{mao10,mao12} showed that there is no coherent vertical field directly above and below the Sun, and \citet{sobey19} found a $2 \kpc$ halo field scale height in the inner Galaxy. In face-on galaxies, a two-layer model of the disk magneto-ionic medium is preferred  \citep{Kierdorf:Mao:2020,Williams:Heald:2024}. However, the interface between the disk and halo is difficult to probe directly. High velocity clouds (HVCs) may offer a way to probe interactions in this interface and gain insight into the magnetic transition.

HVCs trace the cycle of baryons between the disks and halos of star-forming galaxies \citep{Putman:Peek:2012}. Gas falls onto the Galactic disk either as part of a supernova- and stellar wind-driven ``galactic fountain'' process \citep{Bregman:1980} or as fresh infall from outside the Galaxy, falling into the disk with radial velocities (relative to the local standard-of-rest; $\vlsr$) higher than permitted by galactic rotation. As HVCs fall from the halo into the warm disk ($\sim 10^3 - 10^4 \K$; scale height $\sim 1-2 \kpc$ for warm ionized gas [\citealt{Ocker:2020}] and $\sim 100-200 \pc$ for warm neutral gas [\citealt{fost06,malhotra95}]), they interact, forming hydrodynamic instabilities \citep{lock08} and compressing magnetic fields. Evidence of compression of magnetic fields in HVCs has been seen through Faraday rotation of compact extragalactic sources in the Smith Cloud \citep{Hill:Mao:2013,Betti:Hill:2019} and Magellanic Bridge \citep{Kaczmarek:2017}. This effect was first claimed by \citet{McClure-Griffiths:2010} in the Magellanic leading arm, but \citet{Jung:2021} found this detection to in fact be foreground confusion. Simulations also show evidence of compression of magnetic fields as HVCs interact with the disk \citep{Galyardt:Shelton:2016,Jung:2023}.

For this work, we focus on the Anti-Center Shell (ACS; \citealt{Heiles:1984,Tamanaha:1997}), shown in Figure~\ref{fig:hi4pi} with Effelsberg-Bonn \ion{H}{1} Survey (EBHIS) data \citep{wink16}. The ACS is a collection of clouds around Galactic longitude $\ell \sim 180\arcdeg$ separated from Galactic emission by approximately $50 \kms$ in radial velocity \citep{Tamanaha:1997}. The ACS consists of two main filaments on opposite sides of the Galactic plane: a southern filament at Galactic latitude $b \approx -5\arcdeg$ and a northern filament at $b \approx +7\arcdeg$. With previously published observations (all single antenna), the Northern Filament is unresolved and appears as one large, elongated HVC, with its appearance suggesting an extremely large super-shell driven by supernova feedback \citep{Heiles:1984}. Others have argued that it is colliding and merging with the disc underneath \citep{Tamanaha:1997} because it resembles the shapes of simulated HVCs that collide with the disc obliquely \citep{come92}. With $|\vlsr| \gtrsim 80 \kms$ and a separation from the disk gas in velocity space of $\gtrsim 40 \kms$, the ACS meets most definitions of an HVC \citep{Wakker:1991,Putman:Peek:2012}. However, its origin --- stellar feedback or fresh infall --- remains unknown. Whereas the Smith Cloud is relatively high in the halo at height $|z| \approx 3 \kpc$ \citep{lock08} along a complicated sightline passing through the Sagittarius Arm, interacting with the inner Galaxy, the ACS is at $|z| \sim 400 \pc$ in a much simpler direction with little confusion and the large-scale field perpendicular to line of sight.

In this work, we investigate the magnetic field structure in the vicinity of the ACS using new observations from the Dominion Radio Astrophysical Observatory (DRAO) Synthesis Telescope (ST). We describe our observations in Section~\ref{sec:observations}. We then discuss the morphology of \ion{H}{1}-emitting gas and diffuse polarized emission in Section~\ref{sec:morph}. We discuss estimates of the distance to the ACS in Section~\ref{sec:distance}. We discuss the dynamics and kinematics of the region in Section~\ref{sec:dynamics} and the implications for the magnetic field associated with the HVC and the disk-halo region in Section~\ref{sec:fields}. We finish with a summary in Section~\ref{sec:summary}.

\section{DRAO ST Observations} \label{sec:observations}
 
The DRAO ST, located near Penticton, BC, Canada,  consists of seven equatorially-mounted, $\approx 8.6$~m paraboloidal antennas aligned east-west. It  
is described in detail by \citet{land00}.     
We have completed a $\approx 70 \deg^2$ survey of a portion of the northern filament with the DRAO ST, consisting of 62 individual fields. The observations began in 2010 and finished in 2018, with a total of over 1~year of on-sky observing time. We obtained both \ion{H}{1} 21~cm and 
 $1420 \MHz$ continuum observations; these data are shown in 
 Figure~\ref{fig:STmaps}. These new data have substantially improved angular resolution compared to the EBHIS data, resolving the northern filament for the first time. This high resolution, wide-field survey is one of the unique strengths and the hallmark of the DRAO instrument, as is its comparatively complete coverage of $uv$ spacings --- whereas most other interferometer telescopes have incomplete $uv$ coverage and \textit{require} CLEANing, ST maps can benefit from it but, for example, 21~cm \ion{H}{1} line data from the ST are not routinely CLEANed. 
Each of the 62 fields is a synthesis of twelve, 12-hour integrations, with a different configuration of the antennas for each integration.  Baselines from 12.9~m to the maximum 617.1~m are sampled with a regular grid spacing of 4.29~m. Consequently, we have no gaps in $uv$-plane data larger than this, the antenna radius. A single synthesis field has a primary beam with a full-width to half-sensitivity of $107.2'$ at $1420 \MHz$, and resolves structure within the field to a resolution of $49'' \times 49'' \csc \delta \approx 49'' (\mathrm{NS})$, depending on the $uv$-spacing weighting used. This gives an effective $\mathrm{EW}$ beamwidth of $\approx 90'' \times 120''$ for the declination range in our survey area. Structures on the sky from $45'$ down to this resolution limit are captured by the ST alone. The data for individual fields were carefully processed to remove artifacts and to increase the dynamic range using the routines described by \citet{will99}. 
Accurate representation of all structures to the largest scales ($\gtrsim 45'$) is assured by incorporating data observed with large single-antenna telescopes, the John Galt Telescope at DRAO (26~m diameter; \citealt{Wolleben:2006}) and the Effelsberg 100~m telescope \citep{wink16}, after suitable filtering in the Fourier domain. Procedures are described in \citet{tayl03} for the total power (including \ion{H}{1}) data and \citet{land10} for the linear polarization data. In the end, the 62 fields were corrected for the shape of the primary beam, weighted by their relative rms noise and assembled into a large mosaic of the central half of the Northern Filament, covering a rectangle on the sky from $167\fdg5 \le \ell \le 185\arcdeg$ and $4\fdg5 \le b \le 11\fdg5$. The final pixel scale of the mosaic is 18$\arcsec$ per pixel.

\subsection{HI Observations}
The ST produces data cubes centered on the \ion{H}{1} $\lambda=21.1 \cm$ ($1420.4057 \MHz$) spectral line. These data are continuum-subtracted and calibrated to the brightness temperature scale \citep{tayl03}. The final data cube has 256 velocity channels across a $1 \MHz$-wide band, with a velocity resolution $\delta \vlsr = 1.32 \kms$, sampled every $0.82446 \kms$, covering $-164.71 \le \vlsr \le +45.53 \kms$. To increase signal-to-noise sensitivity, we convolve the final data-cube to a resolution of $3\arcmin$. The final 1-$\sigma_\mathrm{rms}$ noise floor in each velocity channel is about 250~mK.

\begin{figure*}[t]
\plotone{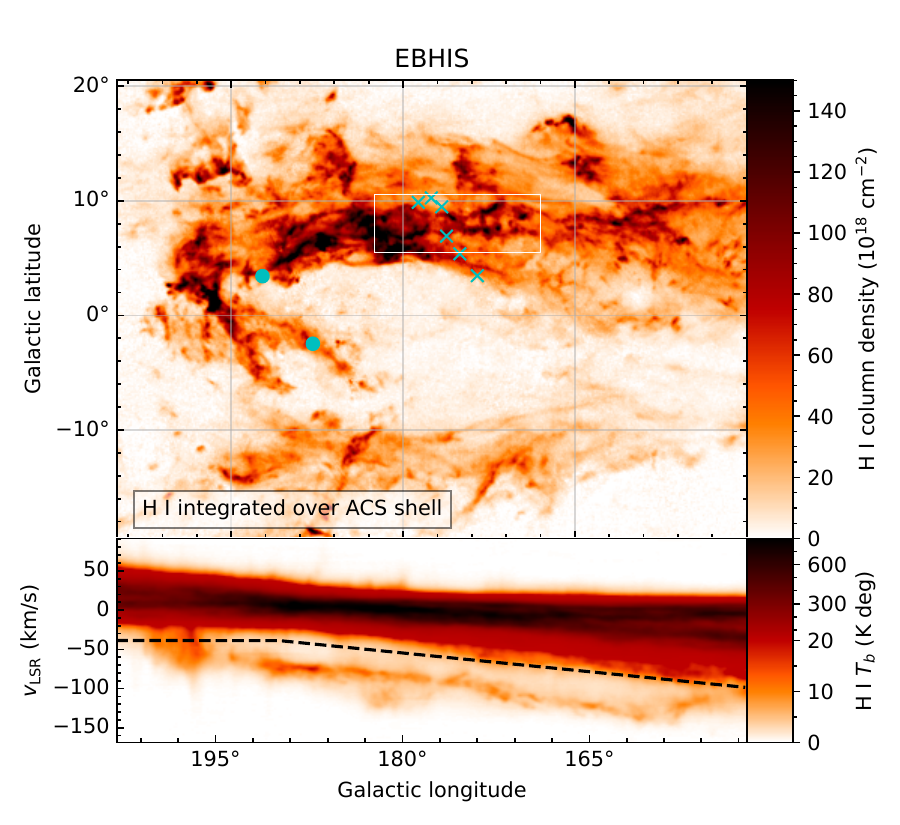}
\caption{Image of the full ACS and corresponding longitude-velocity ($\ell-v$) diagram using EBHIS data. The white box outlines the area observed with the ST, which is shown in Fig.~\ref{fig:STmaps}. The image includes integrated emission at $\vlsr$ below the dashed line in the $\ell-v$ diagram. We show extragalactic compact sources with rotation measure $> 270 \radmsq$ with $\times$ signs and the stars used for distance estimates with dots. The image shows \ion{H}{1} column density derived from the integrated intensity assuming optically thin emission; the $\ell-v$ diagram shows brightness temperature integrated over Galactic latitudes $0\arcdeg < b < +11\arcdeg$. The $\ell-v$ diagram uses a two-slope color scale for intensity.} \label{fig:hi4pi}
\end{figure*}

\subsection{Continuum Observations}
\begin{figure*}[t!]
\plotone{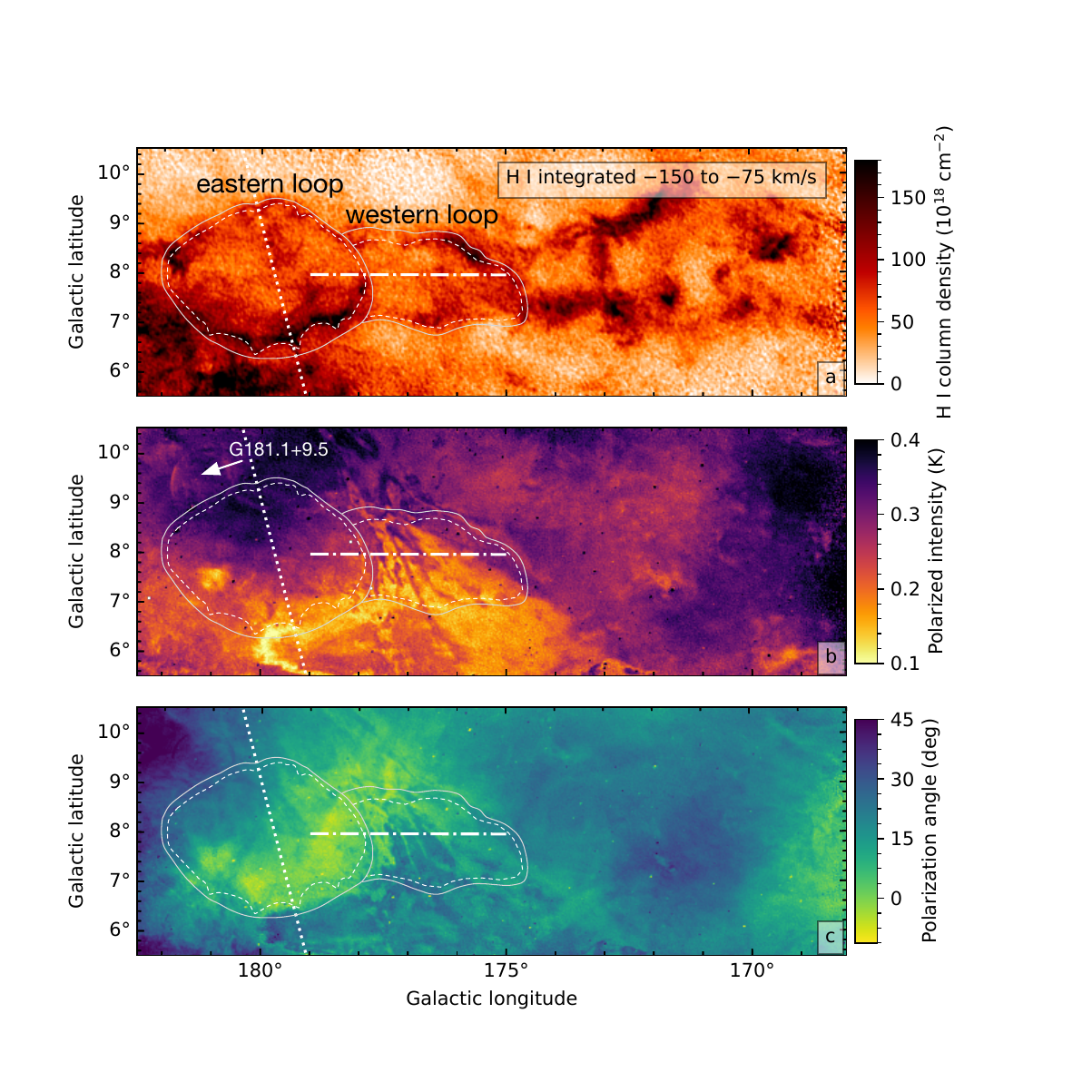}
\caption{Images of the northern filament of the anti-center shell obtained with the DRAO Synthesis Telescope. In panel $a$, we show \ion{H}{1} 21~cm emission integrated over the range of velocities which are distinct from the Galactic plane emission (Fig.~\ref{fig:hi4pi}). With the $\sim 1'$ resolution of the ST, the filament breaks up into a chain of three loop-like structures; we call the two left loops, traced in white dashed lines,  the eastern loop and the western loop. In panel $b$, we show $1420 \MHz$ polarized intensity, in which we can also see SNR G181.1+9.5. In panel $c$, we show polarization angle from the same data. The \ion{H}{1} along the bottom of the eastern loop is especially clearly correlated with a sharp reduction in the polarized intensity and, at the same position, a sharp change in the polarization angle. Between the eastern and western loops is an interface region showing striations of dramatically reduced polarized intensity along with rapid rotations in polarization angle. The dotted and dot-dashed lines show the positions of the profiles in Figs.~\ref{fig:profiles} and \ref{fig:model}, respectively. We interpret the interface as a region with a highly compressed magnetic field resulting in a steep gradient in the rotation measure of this region. This causes the changes in angle and oscillation between constructive and destructive interference, known as depolarization.} \label{fig:STmaps}
\end{figure*}

The ST observes Stokes $I$, $Q$ and $U$ in each of four $7.5 \MHz$-wide continuum bands between $1405 \MHz$ and $1435 \MHz$. Two of the bands are below the \ion{H}{1} line, and the other two are above. We sum the four continuum bands to obtain single $I$, $Q$ and $U$ maps. We follow the standard ST calibration and processing procedure \citep{land10}. From the $Q$ and $U$ maps, we derive maps for polarized intensity, $\PI = \sqrt{Q^2 + U^2}$, and polarization angle, $\psi = 0.5 \arctan(U/Q)$.\footnote{We implemented $\arctan$ using a two-argument inverse tangent function.} Because we do not have sufficient signal to noise in individual channels and we only have EBHIS zero-spacing data at a single frequency, we do not have diffuse emission rotation measures (RMs) and cannot apply any de-rotation of $\psi$. We chose not to correct for the polarization bias, which would only subtract $\sigma \approx 20$~mK in quadrature from \PI ~(the minimum is around 100~mK in Fig~\ref{fig:STmaps}) to minimize modifications to the data.  The continuum flux density is calibrated relative to the \citet{baar77} scale using the compact extra-galactic sources 3C\,147 and 3C\,295; the polarization angle is calibrated with frequent observations of the highly polarized source 3C\,286. The assumed properties of 3C286 at 1420~MHz are a flux density of 14.7~Jy, a fractional linear polarization of $9.2\%$, with an intrinsic polarization angle of {33\degr} and a rotation measure of $-1.2 \radmsq$ \citep{smeg97}. Two levels of correction for instrumental polarization are employed. We use the technique described by \cite{smeg97} to correct for imperfections in the feed and associated waveguide components. We also have to correct for leakage of the Stokes $I$ component into Stokes $Q$ and $U$, which varies with the position across the field of view. \cite{reid08} constructed a correction matrix for amplitude and phase, which was developed with holographic observations of each antenna observing the unpolarized source 3C~295. We used this method for all four individual bands of our continuum observations. 
The symmetrical aperture and well-known polarization characteristics of the ST (and the well-developed processing and calibration procedures above) is another unique strength of the DRAO ST, as it ensures that polarization maps of the sky accurately depict real astronomical structures and minimize instrumental ones.

\subsection{Compact Source Rotation Measure Calculations}
 
Compact, polarized objects observed at 1420 MHz are almost always extra-galactic sources (EGS) with their own structured magnetic field, producing relatively high intensity synchrotron emission ideal for RM calculations. We used the method documented in \citet{vaneck21} to calculate RMs from the linear fit of polarization angle versus $\lambda^2$ of the four closely-spaced bands on the ST. We found RMs for a total of 277 sources in our observations and combined these with the RMs of \citet{vaneck21} to produce Figure~{\ref{fig:RMs_on_PI}}.

\section{Observed morphology}\label{sec:morph}
Here we highlight the unique features of the ACS observations described above. These features are the focus of our investigations.  
\subsection{\ion{H}{1} morphology}

\begin{figure*}[t!]
\centering
\plotone{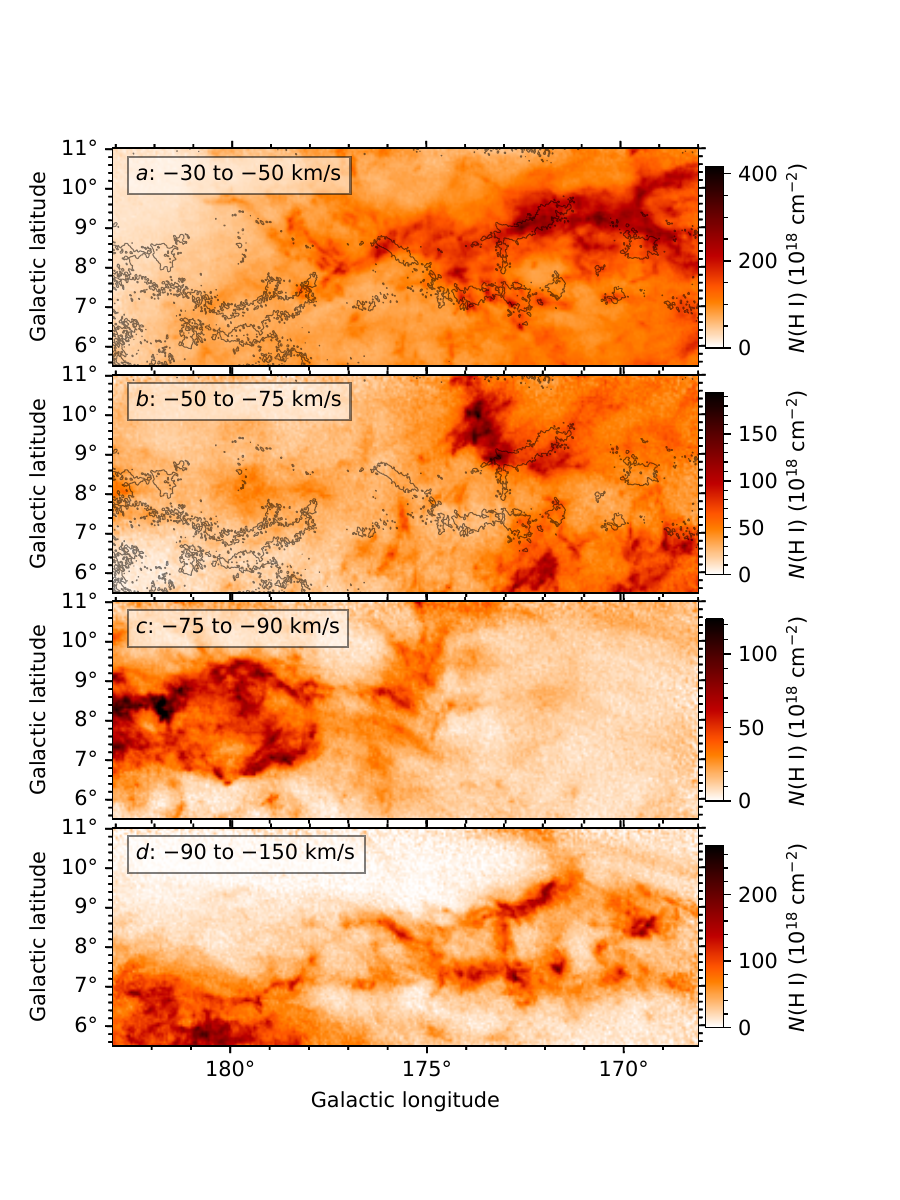}
    \caption{\ion{H}{1} maps integrated over intermediate velocity ranges between the Galactic plane emission and the ACS HVC emission. Contours of the ACS HVC \ion{H}{1} emission at $55 \K \kms$ (Fig.~\ref{fig:STmaps}$a$) are shown in panels $a$ and $b$. The sum of panels $c$ and $d$ is Fig.~\ref{fig:STmaps}$a$.}
    \label{fig:dynamics}
\end{figure*}

In Figure~\ref{fig:hi4pi}, we show the entire ACS region in the EBHIS data. With $10.82'$ angular resolution, the overall morphology is similar to previous maps \citep{Tamanaha:1997}. To the north, there is a filamentary structure around $b=+8\arcdeg$ which appears as a chain of knotted loops of individual HVCs with the chain lying parallel to the plane. In Figure~\ref{fig:STmaps}, we show the new ST map with $1'$ angular resolution in the region outlined with a white box in Figure~\ref{fig:hi4pi}. In the ST \ion{H}{1} data, the loops are the dominant feature. The chain's individual loops have complex, filamentary structures, like strings with beads of \ion{H}{1} clouds. Some clouds have sharp \ion{H}{1} edges. In particular, the sharp edge of \ion{H}{1} around $(\ell, b) = (180\arcdeg, +7\arcdeg)$ in Figure~\ref{fig:STmaps}$a$ suggests shock-compressed gas plowing into the disk from above. At this interface, there are three fingers characteristic of Rayleigh-Taylor instabilities, evident in the \ion{H}{1} along a line from $(\ell, b) \approx (180.0\arcdeg, +6.4\arcdeg)$ to $(178.5\arcdeg, +7.0\arcdeg)$, highlighted by points in the white dashed line tracing the outline of the shell in Figure~\ref{fig:STmaps}$a$. We refer to the loop centred on $(180\arcdeg, +8\arcdeg)$ as the eastern loop and to the smaller one, centred on $\ell = (176.5, +8\arcdeg)$, as the western loop.\footnote{We use north, south, east, and west in Galactic coordinates, with north towards the north Galactic pole ($b=+90\arcdeg$) and east to increasing longitude (the left in a standard map projection).}

In the bottom of Figure~\ref{fig:hi4pi}, we show a longitude-velocity ($\ell$-$v$) diagram of the ACS. The emission from the Galactic plane is evident at low velocities ($|\vlsr| \lesssim 50 \kms$), with positive velocities in the third quadrant ($\ell > 180\arcdeg$) and negative velocities in the second quadrant ($\ell < 180\arcdeg$). The dashed line delineates the planar emission from the ACS emission. The ACS is at large negative velocities, clearly separated in velocity space from the planar emission, with a separation of $\approx 40 \kms$ that is essentially fixed: the ACS emission is parallel to the lowest-velocity planar emission in velocity. 

In Figure~\ref{fig:dynamics}, we show \ion{H}{1} maps in narrow velocity ranges to highlight the kinematic structure of the ACS, providing complementary information to the $\ell$-$v$ diagram. Panels $a$ and $b$ show planar gas which is excluded from the map in Figure~\ref{fig:STmaps}, while panels $c$ and $d$ show gas at ACS velocities which is shown in the map in Figure~\ref{fig:STmaps}. Panels $c$ and $d$ make clear the velocity gradient across the shell, with the eastern loop seen primarily in the $-90 \kms < \vlsr < -75 \kms$ image while the western loop and additional structure to the west seen primarily at $\vlsr < -90 \kms$. However, the small Rayleigh-Taylor finger-like  structures within the compressed gas at the southern edge of the eastern loop are distinct in velocity space from the rest of the eastern loop, being visible only in the $< -90 \kms$ image.

\subsection{Polarization morphology} \label{sec:polmorphology}

\begin{figure}
\centering
\includegraphics[width=0.45\textwidth]{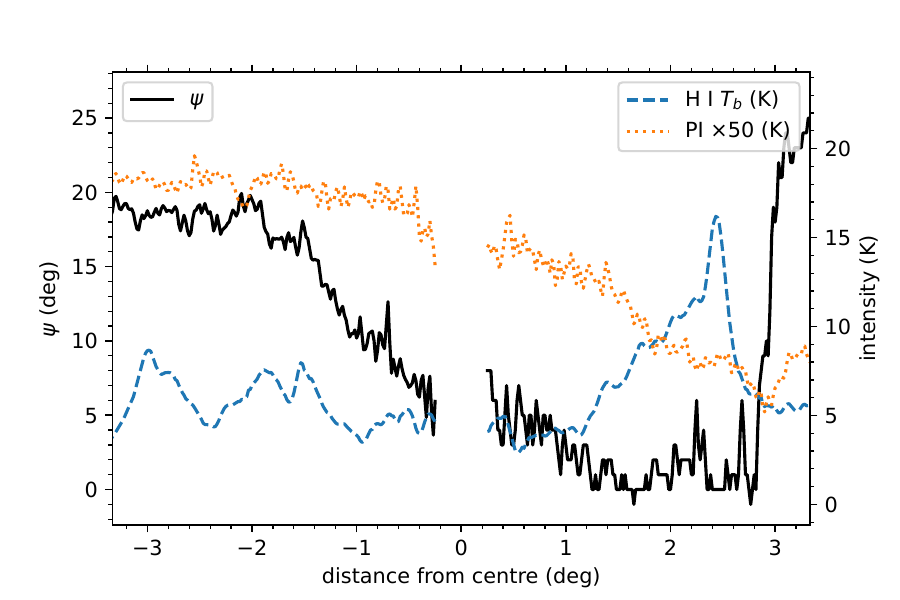}
\caption{Profiles of \ion{H}{1} intensity (blue dashed line; intensity scale on the right), polarized intensity (orange dotted line; intensity scale on the right with a factor of $50$ applied), and polarization angle (solid black line; angle scale on the left) along the diagonal dotted line in Fig.~\ref{fig:STmaps}; positive radius corresponds to southwest (down and to the right in Fig.~\ref{fig:STmaps}). The \ion{H}{1} intensity increases sharply near $2.4\arcdeg$, the front evident in Fig.~\ref{fig:STmaps}. Just outside that front (radius $2.9\arcdeg$), the polarization angle changes by $25\arcdeg$ coincident with a minimum in polarized intensity which we attribute to depolarization.} \label{fig:profiles}
\end{figure}

In Figure~\ref{fig:STmaps}$b$ and $c$, we show polarized intensity (PI) and observed polarization angle ($\psi$) from the ST data. The ACS is part of the so-called Fan Region, an area of bright and smooth PI where $\psi$ changes very slowly both spatially and with frequency \citep{bingham1967,brouw1976}. However within the ACS, there are notable changes in $\psi$. In Figure~\ref{fig:STmaps}$c$, there is a smooth gradient from the top left of the loop, where $\psi \approx 20\arcdeg$, to the bottom right, where $\psi \approx 0\arcdeg$. This is also evident in Figure~\ref{fig:profiles}, in which we profile $\psi$ and PI along the diagonal dotted line in Figure~\ref{fig:STmaps}. Corresponding to this smooth change in $\psi$ is a smooth change in PI, evident in both the image in Figure~\ref{fig:STmaps}$b$ and the profile in Figure~\ref{fig:profiles}. Just below the line of bright \ion{H}{1} on the bottom of the eastern Loop, the polarized intensity reduces dramatically (a dark line in the polarized intensity image and a minimum PI of $\approx 0.1 \K$ in Figure~\ref{fig:profiles}) and the polarization angle changes from $\approx 0\arcdeg$ inside the \ion{H}{1} edge at radius $2.5\arcdeg$ in Figure~\ref{fig:profiles} to $\approx +22\arcdeg$ outside the shell. 


In contrast to the smooth changes within the eastern loop, we observe rapid changes in polarization properties at the interface between the eastern and western loop. In the polarized intensity map, a striated pattern of alternating bright and dark polarization features is seen beginning at $\ell = 178\arcdeg$ between the loops, extending north to south at an angle of $\approx -60\arcdeg$ (counter-clockwise) from the horizontal plane. This is the same as the symmetry axis of the supernova remnant G181.1+9.5, which \citet{koth17} suggested is interacting with the HVC. Concurrent with the striations in polarized intensity are striations in polarization angle. The dashed line at $b=+8\arcdeg$ in Figure~\ref{fig:STmaps} traces a profile through this interface; we discuss and model the morphology of the field in detail in Section~\ref{sec:model} below.

\subsection{Compact source RM morphology}

\begin{figure*}[tb]
\plotone{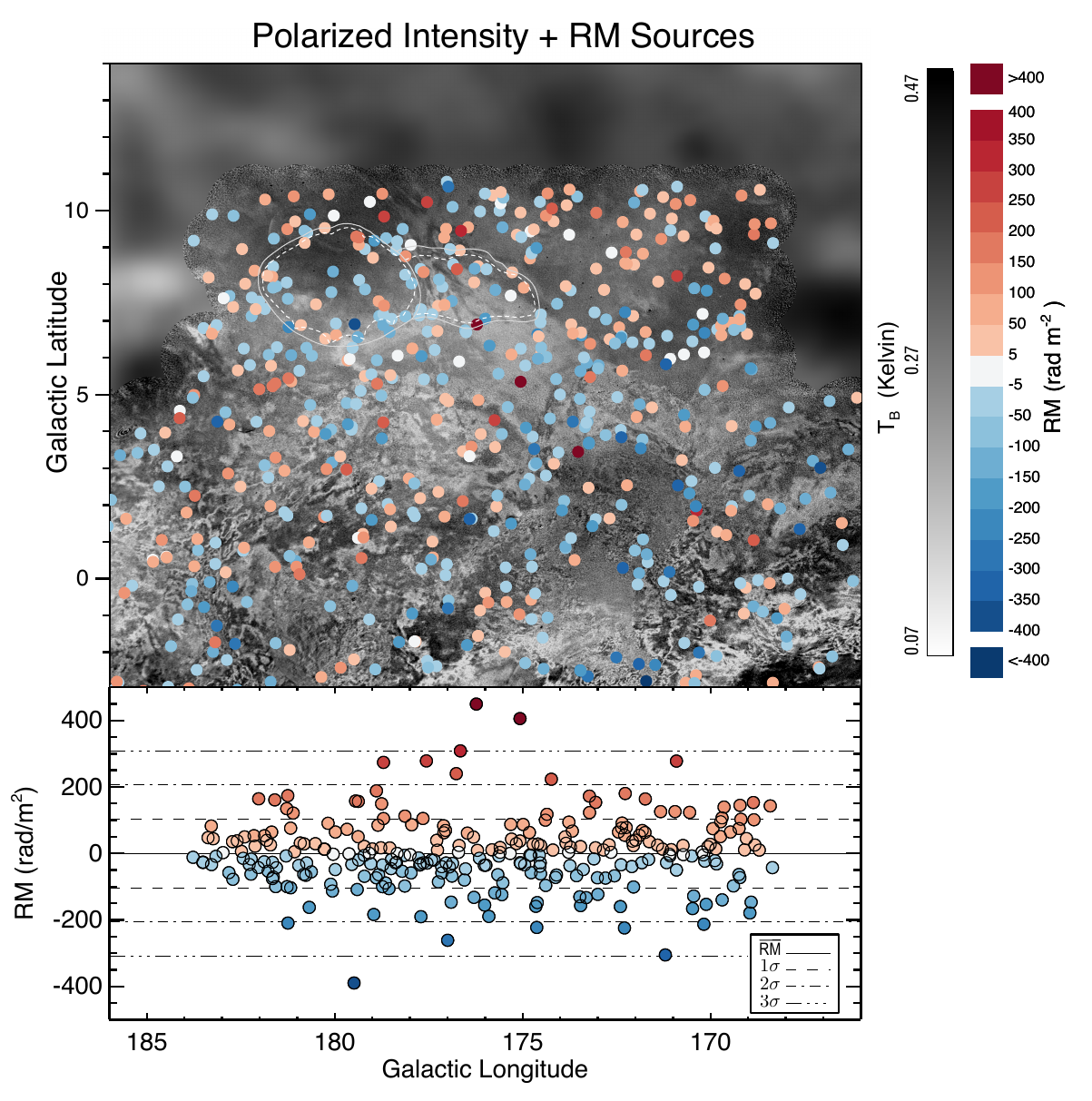}
\caption{Rotation measures of compact sources. Top panel: RMs on a grayscale background of polarized intensity, including our 277 new sources and previously-published CGPS sources; Bottom panel: RMs versus longitude for our 277 sources ($b > 5\arcdeg$). RMs are color-coded into bins as indicated. The solid/dashed shape illustrates the approximate location of the high velocity clouds outlined in Fig.~\ref{fig:STmaps}. The average RM in the top panel is $-18 \pm 115 \radmsq$, while the average RM in the bottom panel is $0 \pm 103 \radmsq$.  The large, positive `ridge' highlighted by 3 RMs $> 400 \radmsq$, plotted in dark red in the middle of the image (coincident with the interface region of Fig.~\ref{fig:sketch}) is beyond $3\sigma$ from the mean. Similarly, there is a large, negative RM source coincident with the bottom of the eastern loop. The six compact sources with $\RM > 270 \radmsq$ are indicated by $\times$ symbols in Fig.~\ref{fig:hi4pi}.}
    \label{fig:RMs_on_PI}
\end{figure*}

As shown in Figure~\ref{fig:RMs_on_PI}, there is a mixture of positive and negative RMs in our region of interest. At low Galactic latitudes ($|b| < 5\arcdeg$), the RMs are predominantly positive at higher longitudes ($180\arcdeg-185\arcdeg$), and predominantly negative at lower longitudes ($170\arcdeg - 175\arcdeg$); at higher Galactic latitudes ($b > 5\arcdeg$), this RM trend is reversed. The magnitudes of the RMs are generally less than $\sim 75 \radmsq$, with the exception of the large, positive `ridge' highlighted by 3 RMs $> 400 \radmsq$, plotted in dark red in the middle of the image. As shown in the lower panel of Figure~\ref{fig:RMs_on_PI}, these large RMs are more than 2-3$\sigma$ from the mean, indicating that these large RMs are not simply the result of the isotropic small-scale field variability.  The six compact sources with $\RM > 270 \radmsq$ are indicated by $\times$ symbols in Fig.~\ref{fig:hi4pi}. These sources appear to align with the striations noted in polarized intensity and polarization angle between the eastern and western shells.

\vspace{3mm}

In the following sections, we analyze the ACS in terms of its distance, motion, and magnetic field structure.

\section{ACS Distance Estimate} \label{sec:distance}

In order to properly study the ACS, we first need to establish its location (distance) in the Milky Way. 
Distances to HVCs are notoriously difficult to constrain. We consider three separate methods that provide consistent evidence for a distance.

The most reliable distance constraints to HVCs in general are based on the presence or absence of interstellar absorption lines at the HVC velocity in spectra of stars with known distances, providing upper or lower limits to the distance to the HVC. \citet{kulk86} reported the absence of Ca~K absorption in 17 stars. These stars have {\em Gaia} parallaxes \citep{GaiaCollaboration:2020} $0.18 \le \pi \le 8.3$~mas. One star in the sample, SAO~78319, is a distance outlier at $\pi = 0.18$~mas; the remaining stars all have $\pi \ge 0.33$~mas, implying a minimum distance to the HVC of 3~kpc. \citet{smoker11} reported no detection of absorption towards HDE~248894 ($\ell, b) = (187.9\arcdeg, -2.5\arcdeg)$ and HDE~256725 ($192.3\arcdeg, +3.4\arcdeg)$, which have {\em Gaia} distances of $3.1 \pm 0.2 \kpc$ and $4.4^{+0.6}_{-0.4} \kpc$, respectively. The positions of these two stars are marked in Figure~\ref{fig:hi4pi}.\footnote{\citet{smoker11} reported these non-detections as a lower limit of $8.0 \kpc$ based on older distance measurements.} However, the absence of absorption could also indicate that a star is behind the HVC but the pencil beam of radiation from the star passes through a hole in the porous HVC; there is no reliable way to rule this out, so we take these lower limits with caution.

A second estimate is obtained by observing that the HVCs are well into their interaction with the disk, and are thus sitting around an ISM scale-height above the midplane. The observed interaction and interface region at $b=+6.5\arcdeg$ implies a distance of $d=8.8 (h / 1 \kpc) \kpc$ for assumed ISM scale height $h$. The scale height for the warm ionized medium (WIM) in the solar neighborhood is $h \approx 1.5 \kpc$ \citep{Savage:2009,Ocker:2020}. The base scale height for \ion{H}{1} in the warm neutral medium (WNM) a range is $h\simeq 200-300 \pc$ \citep[e.g]{malhotra95,fost06}. The thickening or ``flaring'' of the outer Galactic \ion{H}{1} layer means this is a minimum, and $d$ a lower-limit. Due to the sharpness of the \ion{H}{1} shell we take the WNM as the interacting interface, as it is denser than the WIM. There is evidence that the partially-ionized WNM dominates Faraday rotation \citep{fost13,van-eck:2017,Boulanger:2024}. Thus we have a lower-limit estimate of $d\sim 1.8-2.7 \kpc$ to the HVCs.

A third estimate is obtained from the kinematics in Figure~\ref{fig:hi4pi}. While the circular velocity of an object at any distance projected onto the line-of-sight is $0 \kms$ at $\ell = 180\arcdeg$, the gradient of $\vlsr$ with $\ell$ across an extended object is not zero for a non-zero distance. We make the following assumption here: on the large scale, the HVCs are at a common distance and co-moving in their orbit about the Galactic center as one chain. If this is correct the slope along their longitude range entirely reflects their circular galactic rotation. We take the derivative of $\vlsr$ \citep[equation~1 in][]{fost10} with respect to $\ell$, set it equal to the measured gradient across the HVCs, and solve for heliocentric distance $d$. Like the kinematic distance method of solving for $d$ corresponding to a single $\vlsr$, $d$ is very sensitive to the measured $d\vlsr / d\ell$ at $\ell\sim 180\arcdeg$. However, it is easier to measure a systematic gradient in $\vlsr$ here across a longitude range than it is to measure a single systemic $\vlsr$ at a single $\ell$, so this method may yield better kinematic distance estimates along lines-of-sight towards the galactic anti-center. We measure anywhere from 1.2 to $1.4 \kms$ per degree on the $\ell-v$ diagram in Figure~\ref{fig:hi4pi} (across the E and W loops $174\arcdeg\leq \ell \leq 183\arcdeg$), giving an estimated distance range of $3-3.8 \kpc$ with this method \citep[using the solar parameters $R_0=8.122 \kpc$ and $v_0=253 \kms$ from][and the rotation curve, equation~2 in \citealt{fost10}]{Hayes_2018}.

The above three methods return distance estimates based on different assumptions. Methods 2) and 3) overlap at $d \gtrsim 2.7 \kpc$, and the lower limit to the range of 1) is 3.1~kpc, so we adopt $3\pm1.3 \kpc$ as the distance to the ACS, with the substantial uncertainty of $\pm$40\% allowing for the ranges of all three methods. This places the portion of the shell studied with our ST observations $\approx 400 \pc$ above the plane.

\section{Dynamics and kinematics} \label{sec:dynamics}

From the kinematics we note that the eastern and western loops must behave as individual objects, as each is made of \ion{H}{1} seen in its own distinct narrow velocity ranges (Figure~\ref{fig:dynamics}), mostly $-90 \le \vlsr \le -75 \kms$ for the eastern loop and $\vlsr < -90 \kms$ for the western loop. These ranges are separated by somewhat more than the sound speed in the warm ISM, $c_s \approx 10 \kms$. Without magnetic fields, the time since they began interacting must be less than a few sound crossing times because the clouds have not been destroyed. The sound crossing time is $ t_c \sim L /c_s $, where $L$ is the travel length of interest. If we use $L \sim \theta d $, and estimate the angular diameter $\theta$ as $175\arcdeg < \ell < 182\arcdeg = 7\arcdeg$ (i.e., 0.122 rad), the interaction between the cloud and ambient medium must have begun $\lesssim L / c_s \sim (0.122 ~\mathrm{rad})(3 ~\mathrm{kpc}) / (10 \kms) \sim 36 (d/3~\mathrm{kpc})$ Myr ago. Magnetic fields suppress hydrodynamic instabilities, extending the life of the cloud \citep{Jung:2023}. For comparison, the Smith Cloud is expected to collide with the Milky Way in $\sim 30 ~ \mathrm{Myr}$ \citep{lock08}, perhaps as part of an extended stream of gas interacting with the disk \citep{Lockman:Benjamin:2023}.

The $\ell-v$ diagram in Figure~\ref{fig:hi4pi} shows that the HVC emission is separated from the Galactic plane emission by $\approx 40 \kms$. From $\ell = 184\arcdeg$ to $\ell = 168\arcdeg$, the velocity of the HVC emission changes from $\vlsr \approx -80 \kms$ to $\approx -110 \kms$. The limit of Galactic plane emission changes from $\vlsr \approx -30 \kms$ to $-50 \kms$ over the same longitude range, a gradient with the same sign but less steep. Assuming that the filament is co-rotating with the Galactic disk but is falling inward, this is a line-of-sight velocity that reflects mostly a falling motion towards the plane because the typical line-of-sight component of Galactic rotation velocities in this anti-center direction should be negligible. As gas falls into the disc, the Coriolis acceleration, $-2\left(\vec{\Omega} \times \vec{v}^{\prime}\right)$, will give the gas a velocity component in the direction of rotation, towards lower longitudes as viewed from the Sun.\footnote{$\vec{\Omega}$ is the angular velocity vector of Galactic rotation, pointed ``down'' along a direction towards $-z$, and $\vec{v}^{\prime}$ is the velocity of the cloud in the LSR frame, down and towards the Sun.} Therefore, a larger $|\vlsr|$ with the same sign as the gas in the disk is expected. This prograde rotation is also seen in the Smith Cloud \citep{lock08,Lockman:Benjamin:2023}.

The $\ell-v$ diagram in Figure~\ref{fig:hi4pi} and the intermediate-velocity channel maps in Figure~\ref{fig:dynamics} show evidence that the intermediate velocity gas ($-75 \kms \lesssim \vlsr \lesssim -30 \kms$) is interacting with the high-velocity gas ($\vlsr \lesssim -75 \kms$), again similar to behavior \citet{lock08} identified in the Smith Cloud. This is most evident for $\ell < 178\arcdeg$ in Figures~\ref{fig:STmaps} and \ref{fig:dynamics}. The high-velocity loops to the west (right) of the western loop are shown with contours on the intermediate-velocity maps (Figs.~\ref{fig:dynamics}$a$ and $b$). The emission at $\vlsr \approx -40 \kms$, especially the bright filament extending to the right (west) at $(\ell, b) \approx (174\arcdeg, +8\arcdeg)$ is aligned with the high-velocity component, despite the clear separation between high-velocity and low-velocity gas ($|\vlsr| \lesssim 30 \kms$) in the $\ell-v$ diagram and the absence of emission from this filament in the $-50$ to $-75 \kms$ map.

Similarly, an intermediate-velocity loop is evident around $\ell = 176\arcdeg$ to $177\arcdeg$ just above the high-velocity western loop, near a region of rapidly-rotating observed polarization angles that we associate with Faraday rotation (shown in Fig.~\ref{fig:STmaps} and discussed further in Section~\ref{sec:fields}). This supports the idea of compression of colliding clouds as they interact with the disk, also compressing and amplifying the magnetic field. In the $\ell-v$ diagram in Figure~\ref{fig:hi4pi}, there is gas connecting the HVC emission to the disk emission at $\ell \approx 198\arcdeg$, $\ell \approx 185\arcdeg$, $\ell \approx 174\arcdeg$, and $\ell \approx 155\arcdeg$.

\begin{figure*}
   \plotone{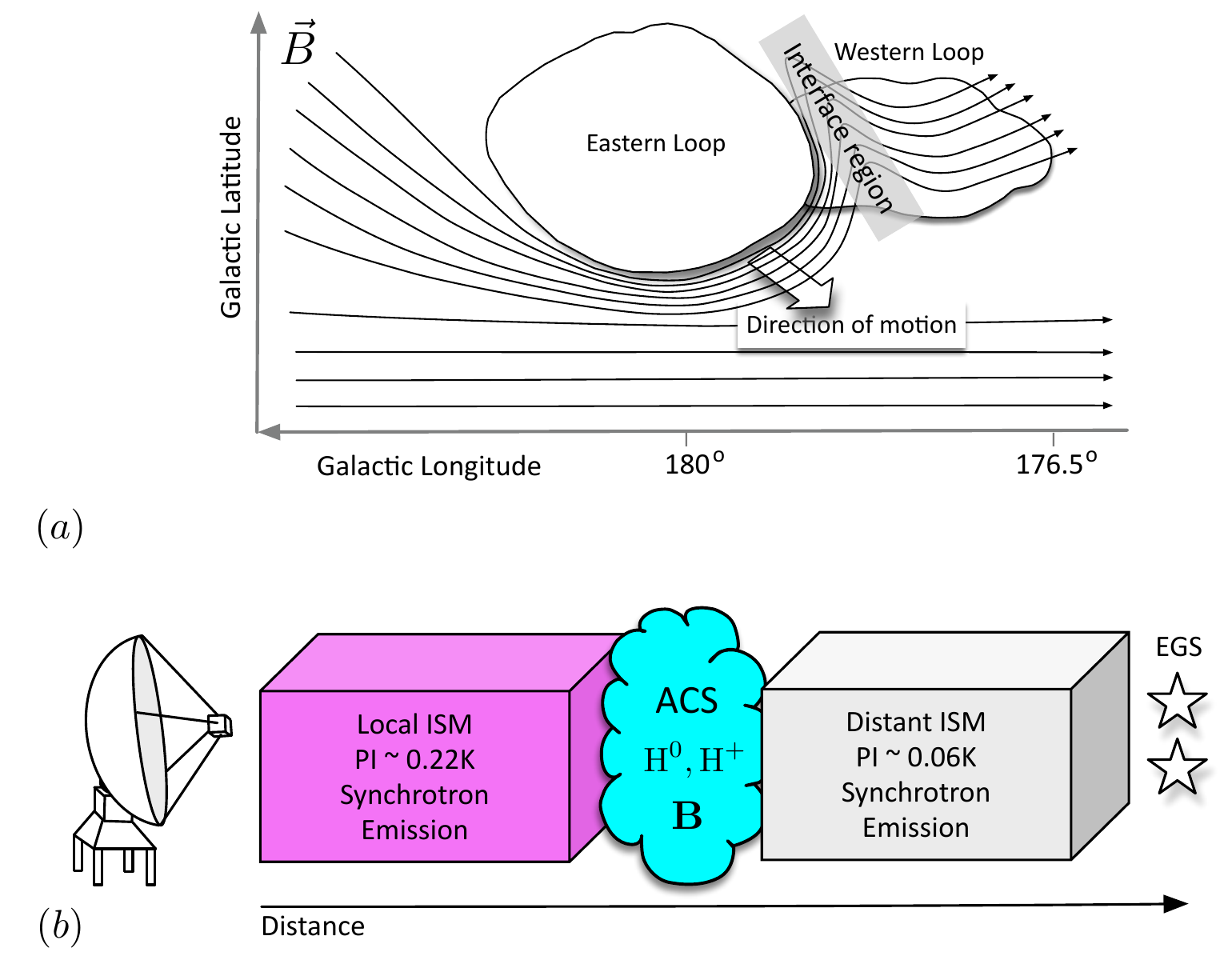} 
    \caption{Sketches of the geometry creating the polarized emission. Panel $a$ shows a possible configuration of the compression of magnetic field lines as the eastern loop collides with the disk and interacts with the western loop, leading to the enhanced rotation measures shown in Fig.~\ref{fig:RMs_on_PI}. Panel $b$ shows the arrangement along the line-of-sight from the telescope to the compact extragalactic sources (EGS).}
    \label{fig:sketch}
\end{figure*}

The morphology of the \ion{H}{1} and polarization features as shown in Fig.~\ref{fig:STmaps} suggests relative motion of the loops, with geometry illustrated in Figure~\ref{fig:sketch}$a$.
We note above three fingers which appear consistent with Rayleigh-Taylor instabilities, instabilities which occur at the interface between two different density fluids. The Rayleigh-Taylor fingers are at higher (more negative) velocities than the bulk of the gas in the eastern Loop (Fig.~\ref{fig:dynamics}$c,d$). These instability fingers should be parallel to the direction of motion on the sky and here they indicate higher-density ACS gas colliding into lower-density gas at the Galactic disk-halo interface. This indicates that the eastern loop is moving to the south-south-west (down and slightly right), as illustrated in Figure~\ref{fig:sketch}$a$.

\section{ACS Magnetic Field} \label{sec:fields}

We use Faraday rotation of diffuse synchrotron radiation and EGSs to determine the magnetic field of the ACS. As the linearly-polarized synchrotron radiation passes through the magneto-ionic gas of the Galaxy, it experiences Faraday rotation by an angle $\Delta \psi = \RM \lambda^2$, where
\begin{equation}
\RM \equiv 0.812 \int_\mathrm{source}^\mathrm{observer} \frac{n_e}{\cucm}  \frac{\mathrm{\bf{B}}}{{\uG}} \cdot \frac{d {\bf{l}}}{\pc} \radmsq.
\label{eq:RM}
\end{equation}
Here $\bf{B}$ is the magnetic field and $n_e$ the electron density along a path element $d\bf{l}$. With the integration direction being from the source to the observer, the dot product between $\bf{B}$ and $d\bf{l}$ means that a positive RM indicates the average $n_e$-weighted, line-of-sight magnetic field is directed towards the observer, while a negative RM indicates it is directed away from the observer.    

\subsection{Ambient Magnetic Field Estimate}\label{sec:ambient}

Just outside the top left edge of the eastern loop we find the supernova remnant (SNR) G181.1+9.5 (indicated in Fig~\ref{fig:STmaps}b), discovered by \citet{koth17}. They showed that this SNR is at the same distance as the ACS, due to morphological association between the HVC's HI clouds and the walls of the SNR. We use G181.1+9.5 to probe the magnetic environment of the ACS using the method of \citet{koth09}. \citet{koth17} were able to determine reliable RMs from diffuse emission at a few positions on the eastern and western walls{\footnote{In studies focused only on SNRs, this would typically be called the `shell' of the SNR. To reduce confusion in this paper, we reserve the word `shell' to refer only to the ACS.}} of SNR G181.1+9.5, by comparing polarimetric observations with the DRAO ST around 1420~MHz and the Effelsberg 100-m radio telescope at 4850~MHz. Those RMs are a combination of foreground Faraday rotation and internal effects. Since the expanding blast wave of the supernova remnant sweeps up and compresses the ambient magneto-ionic medium, the internal magnetic field lines are parallel to the ambient lines in the centre of the SNR's walls \citep{WG1968}. In the center of the eastern wall we measure $\RM = +17.3 \pm 0.8 \radmsq$. As a positive RM, this indicates an average ambient magnetic field pointing towards us. The symmetry axis of bilateral or barrel-shaped SNRs is typically parallel to their ambient magnetic field  \citep[e.g.][]{west16}. This follows naturally from the fact that the expanding blast wave of an SNR is sweeping up the ambient magnetic field and compressing it into its walls. Here we find for the SNR's symmetry axis an angle of $\approx -60\arcdeg$ (counter-clockwise) from the horizontal plane. The same angle we found for the orientation of the polarization features between the two loops. While the eastern wall does not show any significant, systematic change of RM from north to south, the western wall has a small RM gradient from slightly positive in the top to slightly negative in its center. Therefore, the magnetic field is wrapped around the SNR here from the north-east to the south-west. This indicates an ambient magnetic field pointing from the top left to the bottom right \citep{koth09}.

\citet{koth17} determined the ambient particle density around the SNR to be $n_0 \sim 0.6 \cucm (3 \kpc/d)$, scaled to our new distance $d$ of 3~kpc to the SNR and the ACS. 
This translates to an average electron density $n_e \sim 5.3 \cucm (3 \kpc / d)$ inside the SNR walls for a fully ionized plasma with a compression ratio of 8 for the wall of the SNR  
and a gas of $10\%$ helium and $90\%$ hydrogen by number of atoms.

Large-scale Galactic magnetic field models fit to RMs of extragalactic sources and pulsars indicate that the magnetic field in the outer Galaxy is roughly perpendicular to the line-of-sight (e.g., \citealt{vaneck11}).  Therefore, we can assume that the contribution to the observed RM from the Milky Way at these longitudes is essentially zero. By assuming a spherical SNR, we can calculate the maximum effective path length $\Delta l$ through the wall of the SNR on the eastern side (i.e., where there is an observed radial peak in PI as seen in Fig.~\ref{fig:STmaps}$b$). The longest line of sight through the wall is at the inner edge of the wall and can be geometrically calculated as $13 \pc (d/3 \kpc)$. 

Furthermore, if we assume that inside the SNR wall, the synchrotron-emitting material is mixed with the Faraday-rotating plasma, then we would expect that the observed RMs for this extended emission region should be half of what would be observed for an extra-galactic compact source along the same line-of-sight (i.e.,  the so-called ``Burn slab'' model; \citealt{burn66}).  By including the factor of 1/2 from the Burn slab model in equation~(\ref{eq:RM}),  we can approximate the observed RM for the SNR wall as
\begin{equation}
\RM_{\rm{wall}} \sim \left(\frac{1}{2}\right) 0.812 \langle n_e \rangle \langle B_{||} \rangle \Delta l.
\label{eq:RMwall}
\end{equation}
Rearranging this, we can then estimate the {\it{compressed}}  magnetic field along the line of sight inside the wall as $B_{||}$, using the values for $n_e$ and $dl$ above, to find
\begin{eqnarray} \label{eq:B}
B_{{||\rm{wall}}}&=&  +0.62 \uG \; \left(\frac{\RM}{17.3 \radmsq}\right).
\end{eqnarray}
 This result is entirely independent of the distance to the ACS, since the distance dependence of our estimates for $n_e$ and $\Delta l$ cancel out.  Correcting for the magnetic field compression \citep[e.g.,][]{WG1968}, this implies the ambient $B_{||}$ is actually only $\approx 0.05\uG$. This confirms the idea that the ambient magnetic field is essentially perpendicular to the line-of-sight in this direction, not only in the plane of the Galaxy, but also in the environment of the ACS. 

\subsection{Compressed Fields in the Interface Region}

The rapid changes in PI and $\psi$ in the striated interface between the eastern and western loops described in Section~\ref{sec:polmorphology} suggest that the complex polarization vector $\mathcal{P} \equiv Q + iU$ associated with that physical structure rapidly rotates through $180\arcdeg$ causing, alternatively, constructive and deconstructive interference (and depolarization). We can readily explain the striations described in Section~\ref{sec:polmorphology} with Faraday rotation, a result of ionized gas and magnetic fields rotating the polarization angle of background polarized emission. We illustrate the geometry in Figure~\ref{fig:sketch}.

In our picture, the eastern loop is interacting with the western loop, creating a compressed magneto-ionic medium in the interface region around $\ell = 177\arcdeg$, while simultaneously compressing the magnetic field as it falls into the disk. The compression process between the two clouds twists the magnetic field from being in the plane of the sky to having a component along the line of sight which causes enhanced Faraday rotation (Fig.~\ref{fig:sketch}$a$), as seen in the large RMs of extragalactic compact sources in this region (Fig.~\ref{fig:RMs_on_PI}). Additionally, the compressed magnetic field at the bottom of the eastern loop causes similar enhancements to extragalactic RMs. Emission from the distant ISM is Faraday-rotated by the medium in the interface, while emission from the foreground is not (Fig.~\ref{fig:sketch}$b$). Therefore, we observe the superposition of the heavily Faraday-rotated background (emitted in the gray-shaded area and Faraday-rotated first by the cyan-shaded ACS and then by the magenta-shaded area in Figure~\ref{fig:sketch}$b$) and the lightly Faraday-rotated foreground (both emitted and Faraday-rotated in the magenta-shaded region). With only the four frequency channels of the ST, we have no ability to use Faraday synthesis \citep{brentjens:2005} to directly disentangle multiple components along the line of sight. Therefore, the repeated rotations through $180\arcdeg$ of the background emission by the ACS produce much smaller oscillations of the observed $\psi$ because the majority of the PI is produced in the foreground, which is rotated much less.
 
We model this as a gradient in electron density from east to west (left to right) across the compressed interface region, attempting to explain the data in a profile at $b=+8\arcdeg$ shown as dotted lines in Figure~\ref{fig:model}. As the compressed interface region rotates the background polarized emission, the background is first rotated such that it is maximally depolarized, around $\ell = 177\fdg 5$. Then further rotation leads to constructive interference around $\ell = 177\fdg 3$, and yet more rotation leads again to destructive interference around $\ell = 176\fdg 7$. The observed polarized intensity and polarization angle return to close to the background levels around $\ell = 176\fdg5 4$, at the edge of the interface.

\subsection{ACS Magnetic Field Toy Model} \label{sec:model}

\begin{figure}
\includegraphics[width=0.47\textwidth]{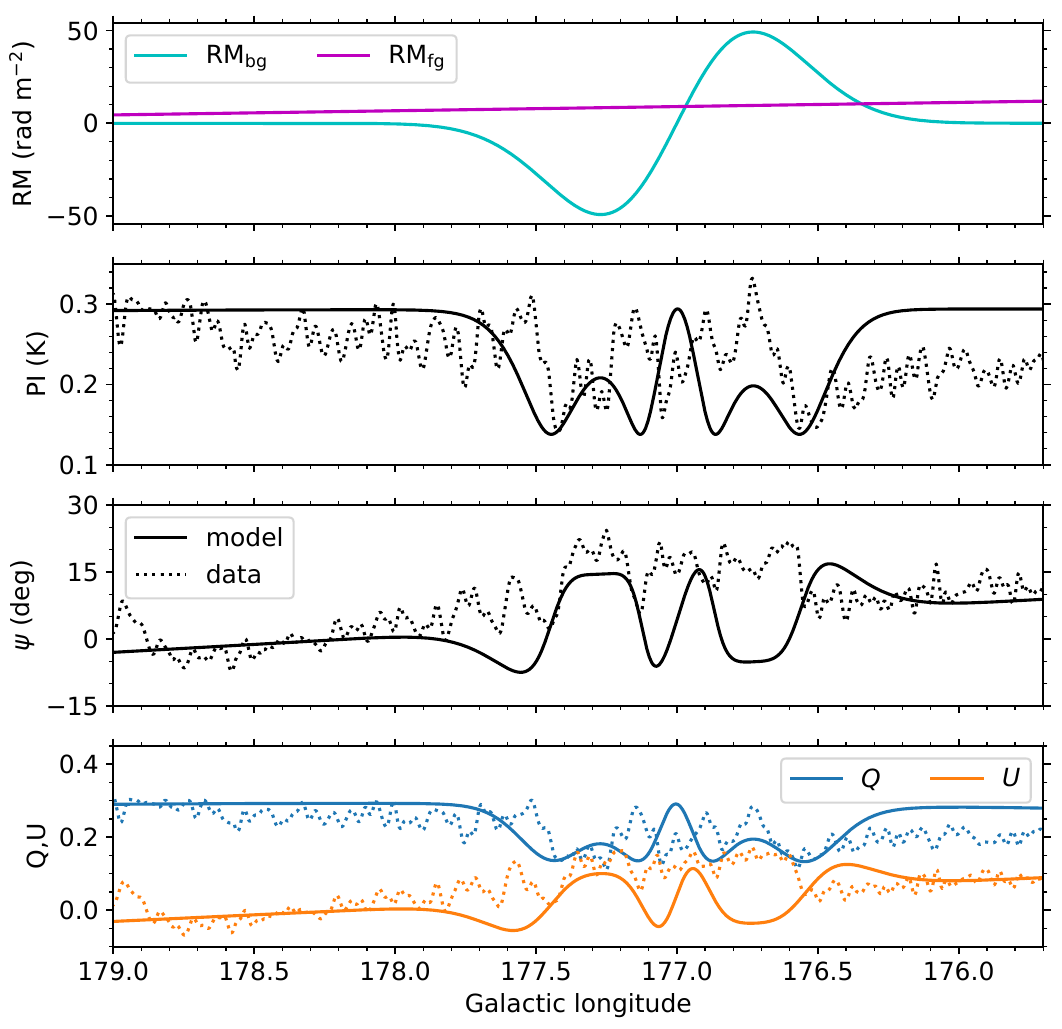}
\caption{Results from the toy model described in \S~\ref{sec:model}. Top panel shows the modeled RM contributions from the shell (cyan) and the Milky Way (MW; magenta). The next three panels show the PI, polarization angle $\psi$, and Stokes $Q$ (blue) and $U$ (orange) from the model (solid lines) compared to our data (dotted lines) at $b=+8.0\arcdeg$ (the dot-dashed line in Figure~\ref{fig:STmaps}).}
\label{fig:model}
\end{figure}

We developed a toy model to simulate the observed characteristics in polarized emission and depolarization across the interface region, quantifying the discussion above. The data we model are shown as dotted lines in Figure~\ref{fig:model}, which are taken from the horizontal dashed line at $b=8.0\arcdeg$ in Figure~\ref{fig:STmaps}.  

Key characteristics of the observations that we aim to address are as follows. From $\ell \approx 179\arcdeg$ to $177.8\arcdeg$, the measured PI is relatively constant at $\approx 0.28 \K$ while the polarization angle increases from $\psi \approx -5\arcdeg$ to $\approx +10\arcdeg$. Equivalently, some power switches from $Q$ to $U$. Starting at the left (eastern) edge of the polarization feature around $\ell = 177\fdg 8$, we see PI drop sharply to a minimum of $0.16 \K$, indicating that the polarized background emission has been rotated $90\arcdeg$ relative to the polarized foreground emission, resulting in depth depolarization.  Mathematically, this can be described as having $\mathcal{P}$ fully cancelling with the foreground emission vector.

If the change in polarization is due to changing Faraday rotation of the background emission by the ACS, the background PI should be half the difference between the PI outside the depolarizing feature (ie $0.28 \K$)  and the minimum ($0.16 \K$), giving us a background PI of $0.06 \K$ and a foreground PI of $0.22 \K$. This implies that $\sim 20\%$ of the synchrotron emission originates behind the ACS. This is consistent with estimates in the Fan Region that some emission originates $\gtrsim 2 \kpc$ away beyond the Perseus Arm \citep{Hill:2017} while most originates within $\lesssim 500 \pc$ \citep{Iacobelli:2013,West:Landecker:2021}.

At local maxima, the foreground and background polarized emission constructively interfere, whereas at local minima, the foreground and background destructively interfere. Continuing westward, the background is further rotated and PI increases as the difference in angle between the background and foreground emission decreases. We observe a maximum PI at $\ell \approx 177\fdg 2$, near the center of the interface between the eastern and western loops, at which point the background emission is rotated by about $180\arcdeg$. Because we observe the sum of the background polarization vector and the foreground polarization vector, we observe a total rotation of only about $25\arcdeg$; the amplitude of the observed rotation provides information about the contribution of the foreground and background polarized emission. 

At an observing frequency of $1420 \MHz$, each rotation of the background $\mathcal{P}$ by $\pi$~radians requires a change in the RM of $\Delta \RM = \pi / \lambda^2 \approx 71 \radmsq$. This implies a change in the magnetic field in the ACS of
\begin{equation}
B_{||} \approx 5.5 \uG \left( \frac{n_e}{5.3 \cucm} \right) \left( \frac{d}{3 \kpc} \right)
\end{equation}
per rotation (using equation~(\ref{eq:RMwall}), but without the factor of 1/2 because the rotating and emitting regions are not assumed to be mixed).

\newcommand{\shell}{\ensuremath{_{\mathrm{s}}}}
\newcommand{\bg}{\ensuremath{_\mathrm{bg}}}
\newcommand{\fg}{\ensuremath{_\mathrm{fg}}}
\newcommand{\Pol}{\ensuremath{\mathcal{P}}}
\newcommand{\Pbg}{\ensuremath{\Pol\bg'}}
\newcommand{\Pfg}{\ensuremath{\Pol\fg'}}
\vspace{5mm}

Our model of this emission is constructed out of four components: two polarized emitting regions (foreground and background) and two Faraday-rotating regions (the ACS and the foreground). 
We assume foreground $\mathrm{PI}=0.22 \K$ and background $\mathrm{PI} = 0.06 \K$ as estimated above, and as illustrated in Figure~\ref{fig:sketch}$b$. We split the power between $Q$ and $U$ as $\Pfg = 0.22 - 0.04i$ and $\Pbg = 0.06 -0.01i$ to match the observed polarization angles. We then apply a constant Milky Way RM of $0$ at $\ell = 179\arcdeg$ with a linear increase to $18 \radmsq$ at $\ell = 175\arcdeg$; this RM rotates both the foreground and background emission. It is the derivative of RM (i.e., $d\RM / d\ell$) that is constrained by the observed slope in $Q$ and $U$; we have freedom to adjust the value of the \RM\ at the reference point ($\ell = 179\arcdeg$) by also adjusting the power split between $Q$ and $U$ in \Pbg\ and \Pfg. We choose a slightly positive value of \RM\ because in the background $\psi \approx 0\arcdeg$ and in the shell $\psi \approx +20\arcdeg$.

We assume two regions of polarized radio emission, producing $\Pbg$ from the background and  $\Pfg$ from the foreground. We observe
\begin{equation}
\Pol = \Pfg e^{i \, \RM\fg \, \lambda^2} + \Pbg e^{2 \, i \, \left( \RM\bg + \RM\fg \right) \lambda^2}
\end{equation}
after the polarized radiation propagates through the RM of the shell ($\RM\bg$) and the RM of the foreground ($\RM\fg$).  As discussed above, in regions where the emission and rotation regions are smooth and mixed (i.e., a \citealt{burn66} slab), the \RM\ of the emission is half that which would be observed for extragalactic compact sources, so there is no factor of two in the first term. For $\RM\bg$, we assume a Gaussian shape multiplied by a sine function to change the sign as suggested by our assumed field geometry (Fig.~\ref{fig:sketch}):
\begin{equation}
\RM\bg = \RM\shell \sin \left( \frac{\ell - \ell\shell}{w} \right) \exp \left( -\frac{(\ell - \ell\shell)^2}{\sigma^2} \right).
\end{equation}
This is the cyan line in the top panel of Figure~\ref{fig:model}, with $\ell\shell = 177\arcdeg$, standard deviation $\sigma$, $\RM\shell = -300 \radmsq$, and characteristic width $w=1\arcdeg$. 

We ran a suite of realizations of this toy model, identifying the values described above as plausible (but not unique) matches for the data.
Of the free parameters in the model, $\RM\shell$ affects the period of the oscillations in $Q$ and $U$ (and therefore PI and PA), while the fraction of the emission originating in the background affects the amplitude of the oscillations. If we assume the magnetic field is the compressed ambient field observed around SNR G181.1+9.5, we can use equation~\ref{eq:B} to estimate a line-of-sight field $B_{||} \approx 5 \uG$. The negative sign in $\RM\shell$ determines the sense of the rotation, such that the first large peak (at $\ell = 177.3\arcdeg$) has increased $\psi$, whereas it would be in the opposite phase if we chose $\RM\shell = +300 \radmsq$. We chose a Gaussian times sine shape of the enhanced RM for mathematical simplicity to illustrate the model but do not suggest that this describes the true shape of the enhanced RM; more detailed modeling of the compression of the field between the shells is beyond the scope of this paper. However, this simple model is sufficient to illustrate how a narrow region of enhanced RM produces the kind of change in PI and $\psi$ that we observe across the ridge between the eastern and western shells.

\subsection{Extra-Galactic Source RM Analysis} \label{sec:RMs}

As discussed in section~\ref{sec:morph}, at low latitudes, the RMs in this region are predominantly positive at higher longitudes, and then switch to being predominately negative at lower longitudes. This is expected in the Galactic disk with the large-scale field being very nearly perpendicular to the line of sight in this longitude range, and directed clockwise as viewed from the North Galactic pole \citep{vaneck11}. In contrast, at latitudes higher than ($b > 5\arcdeg$), this pattern is reversed, as noted by \citet{Dickey:2022}. Figure~\ref{fig:sketch}$a$ reflects this idea with the magnetic field lines being drawn in opposite directions between the lower and higher latitudes. 

We attribute the enhanced RMs in the interaction region between the eastern and western loops to be the result of  magnetic field compression and amplification as the loops fall into the Galactic disk, dragging the magnetic field with them. These actions create an increase in the line-of-sight component of the magnetic field sufficient to create large RMs. We have sketched this interaction region (i.e., the low halo region)  as being largely co-planar with the disk field.  We stress that the actual geometry is not limited to this; the key is that the line-of-sight component is being affected by the motion and interaction of the loops in this region, with the halo field being compressed and drawn into the disk by the HVCs. 

We also note that the large RMs ($>400 \radmsq$) extend well below our interacting eastern and western loops. This may be an indication that the lower section of the Northern filament, as shown in Figure~\ref{fig:hi4pi}, is also contributing to the compression of the magnetic field in this region through similar motion to these loops.

\section{Summary} \label{sec:summary}

We have shown that the ACS, first identified as a single elongated HVC in single-antenna observations, is actually composed of smaller loop-like individual HVCs. While the delineation of the individual clouds can be seen upon closer inspection in the EBHIS data, the separate loops are clear in our new ST observations. The eastern and western loops have unique and distinct velocity profiles, indicating they must be separate objects. 

Additionally, we see evidence that these two loops are falling onto the Galactic disk, while simultaneously interacting with each other. The falling motion of the loops is revealed by Raleigh-Taylor ‘fingers’ at the bottom of the eastern loop, as the higher-density ACS gas collides into the lower-density gas of the Galactic disk-halo interface. We also see a polarization angle jump that implies that the ambient magnetic field has been swept up by the eastern loop and is now wrapped around this HVC in a thin layer of electrons. This ionized gas was produced by friction while the HVC is plowing into the Galactic disk at the disk-halo interface.

Through an analysis of observations of the nearby SNR G181.1+9.5, which \citet{koth17} argued is at the same distance as the ACS, we showed that the line-of-sight ambient magnetic field in this region is less than 0.05~$\uG$ – meaning that the large-scale field is essentially perpendicular to our line of sight. This is consistent with the conclusions of previous models of the Galactic magnetic field. It then follows that the observations of the enhanced RMs at the interface of the two loops is the result of the magneto-ionic medium being compressed and twisted, resulting in a significant line-of-sight component of the magnetic field, thereby producing the large RMs. 
 
Our result is the first measurement of magnetic fields associated with an HVC in polarized emission, and the first with a map that clearly illustrates the magnetic field geometry. 
In the future, wide-field maps of the Milky Way's mid-latitudes made with interferometers with outstanding polarization fidelity, wide fields of view, and wide bandwidths to enable Faraday synthesis such as the Australian Square Kilometre Array Pathfinder \citep{Thomson:2023}, the upgraded DRAO ST \citep{Landecker:Belostoski:2019}, and the Square Kilometre Array itself \citep{Dewdney:2009}, will be able to further exploit HVCs as tools for visualizing the magnetic field in the disk-halo interface.

\begin{acknowledgements}
TJF and BF thank the Brandon University Research Committee for a 2018 Research Grant. The authors also gratefully acknowledge the support of DRAO, a National Facility operated by the National Research Council Canada, and their commitment to observing all the fields necessary for this work, particularly Operations Manager Andrew Gray for his diligence in scheduling them. ASH was partly supported by the Dunlap Institute at the University of Toronto and an NSERC Discovery Grant. We acknowledge useful discussions with T. Landecker and A. Ordog. ASH acknowledges useful discussions regarding this paper at the Interstellar Institute, a CNRS International Research Network, especially with R.~A.~Benjamin. We also thank the anonymous referee for a careful reading which led to an improved paper.
\end{acknowledgements}

The datasets generated and/or analysed in this study are available from the corresponding author upon reasonable request.

\bibliography{references}

\end{document}